\documentclass[aps,prc,preprint,superscriptaddress,showpacs,floatfix]{revtex4}

\usepackage{CJK}
\usepackage{graphicx}

\begin{document}

\title{Description of the critical point symmetry in 124Te by IBM-2}

\begin{CJK*}{GBK}{song}

\author{DaLi Zhang}
\email[Corresponding author:] {zdl@zjhu.edu.cn}
\affiliation{Department of physics, Huzhou University,
Huzhou 313000, Zhejiang, China}
\author{ChengFu Mu}
\affiliation{Department of physics, Huzhou University,
Huzhou 313000, Zhejiang, China}

\date{\today}

\begin{abstract}
Based on the neutron and proton degrees of freedom, the low-lying energy levels, the $E2$, $M1$ and $E0$ transition
strengths of nucleus $^{124}$Te have been calculated by the neutron-proton interacting boson model. The calculated
results are quite consistent with the experimental data. By comparing the key observables of
states at the critical point of $\textrm{U}_{\pi \nu}(5)$--$\textrm{O}_{\pi \nu}(6)$ transition
with experimental data and calculated results, we show that the $^{124}$Te is a possible nucleus
at the critical point of the second order phase transition from vibration to unstable rotation and
such a critical point contains somewhat triaxial rotation. The 0$^+_2$ state of $^{124}$Te
can be interpreted as the lowest state of the first excited family of intrinsic levels in
the critical point symmetry.\\
Key words:  $^{124}$Te, low-lying structure, the critical point,
$\textrm{U}_{\pi \nu}(5)$--$\textrm{O}_{\pi \nu}(6)$
transition.\
\end{abstract}

\pacs{21.10.Re, 21.60.Fw, 27.60.+j}

\maketitle

\end{CJK*}


\section{Introduction}
\label{s1}

How to characterize the underlying structure of $^{124}$Te is a difficult task and has
attracted considerable interests in variety of theoretical and
experimental studies \cite{Garrett18,He17,Hicks17,Garrett16,Sabria16,Saxena14}. The $^{124}$Te lies outside the
major shell $Z=50$ with two protons, and its ratio $E(4^+_1/2^+_1)=2.07$ is very close to the
spherical vibrator value of 2.00. This indicates that the $^{124}$Te is a vibrational or U$(5)$
symmetric nucleus. However, the energy levels spread of the two-phonon multiplets do not
suggest a vibrational character in $^{124}$Te. Meanwhile, the energy
level of the lowest excited 0$^+$ state suggests that the $^{124}$Te exhibits more $\gamma$-unstable
or O$(6)$ behavior, but the $E2$ decays of 0$^+_2$ state do not support the  $\gamma$-soft
rotational nature. \

On the other hand, shape coexistence in the Te isotopes is still elusive \cite{Garrett16}.
The appearance of strong $\rho(E0, 0^+_2\rightarrow0^+_1)$ transition is an important spectroscopic
fingerprint that mostly supports the shape coexistence in nucleus \cite{Heyde11}.
For $^{124}$Te, the experimental value of the electric monopole transition strength
$\rho^2(E0, 0^+_2\rightarrow0^+_1)\times10^3$ is $12\pm3$ \cite{T05}. This value is very similar
to the corresponding transition in the Cd isotopes, the latter demonstrates the firm evidence for the
nature of shape coexistence, leading to strong support for deformed intruder structure of the first excited 0$^+$ state
in $^{124}$Te isotopes, but not providing conclusive proof for shape coexistence in the $^{124}$Te yet \cite{Garrett16}.
Meanwhile, the systematic research on the known experimental information suggested that $^{124}$Te may be
an example of nucleus at the $E(5)$ critical point \cite{Clark04}. The recent investigation
indicates that $^{124}$Te exhibits $E(5)$-like structure \cite{Sabria16, Ghit08}.\
But there are some debates on understanding the low-lying structure of the $^{124}$Te
( see, e.g., Refs.\cite{Hicks17, Ghit08, Guliyev02} and references therein).\

The $E(5)$ and $X(5)$ symmetries originally developed by Iachello \cite{Iachello00,Iachello01}
correspond to an exact solution of the Bohr Hamiltonian for $\gamma$-independent potentials
and an approximate solution of the Bohr Hamiltonian for $\gamma\approx 0^\circ$, respectively.
From the Bohr Hamiltonian with $\gamma$ frozen at $\gamma=30^\circ$, a four dimensional
critical point symmetry model called $Z(4)$ has been introduced \cite{Bonatsos2005}. Most
recently, another four dimensional critical point symmetry called $T(4)$ is obtained from
the Bohr Hamiltonian with the $\beta$-soft potential and for a fixed value of $\gamma$ with
$0^\circ \leq \gamma \leq 30^\circ$ \cite{Zhangyu2017}.\

In nuclear system, $E(5)$ symmetry can be used to describe a nucleus at the critical point of the
second order phase transition from a spherical vibrator to a $\gamma$-soft rotor.
In the interacting boson model (IBM-1) \cite{Iachello87}, which do not distinguish the proton boson and neutron boson,
the $E(5)$ represents a nucleus that is located at the critical point of the shape phase transition
between the U$(5)$ and O$(6)$ symmetries. Since the neutron and proton degrees of freedom are explicitly
taken into account, the proton-neutron interacting boson model (IBM-2)\cite{Iachello87} possesses  a
complex phase diagram \cite{Caprio05, Cejnar09}. The critical point of second order phase transition
can occur at the phase transition from $\textrm{U}_{\pi \nu}(5)$ to $\textrm{O}_{\pi \nu}(6)$ symmetry,
and can be subsumed to the critical point of the second order phase transition in IBM-1 \cite{Caprio05, Cejnar09, Cejnar10}.
Compared with their neighboring isotopes, nuclei at or close to the critical point of shape phase transition
demonstrate dramatic changes in the properties of the low-lying states, such as the energy levels,
the $E2$ transition strengths and $E2$ branching ratios \cite{Naz18}.
But there is no discussion on the critical point symmetry in the real nuclei for the phase transition
between $\textrm{U}_{\pi \nu}(5)$ and $\textrm{O}_{\pi \nu}(6)$ in the IBM-2 phase diagram, so far.\

The low-lying levels of $^{124}$Te have recently been examined with the
(${n,n',\gamma}$) reaction. The spins, level energies, $B(E2)$ transition probabilities,
and multipole-mixing ratios were obtained \cite{Hicks17}. The experiment provides us an opportunity to
deeply understand the low-lying structure of $^{124}$Te. In this work, we study the
low-lying structure of the $^{124}$Te within the framework of the IBM-2, with special attention
paid to the key observables of the critical point of the second order transition between
$\textrm{U}_{\pi \nu}(5)$ and $\textrm{O}_{\pi \nu}(6)$. We also calculate the low-lying energy
levels, the $E2$, $M1$ and $E0$ transition strengths and compare the predications of the critical
point symmetry of the $\textrm{U}_{\pi \nu}(5)$--$\textrm{O}_{\pi \nu}(6)$ transition with
the experimental data. We attempt to describe the critical point symmetry in the $^{124}$Te,
and reveal the low-lying structure of $^{124}$Te in IBM-2 space.\

The outline of the paper is as follows. In Sec.~\textrm{\ref{s-2}}, we introduce the Hamiltonians
of IBM-2, $E2$, $M1$, and $E0$ operators used in this work. The criteria
adopted for the determination of the model parameters, and the comparison of the
numerical results with experimental data and the predications of the critical point symmetry
are presented in Sec.~\textbf{\ref{s3}}. Finally, out concluding remarks and summary
are stated in Sec.~\textrm{\ref{s4}}.\

\section{Theoretical framework}
\label{s-2}
The IBM-2 is an algebraic model, in which valence nucleons are coupled to
form $s_{\rho}$ bosons (angular momentum $L$=0) and $d_{\rho}$ bosons (angular momentum $L=2$) with $\rho= \pi$
standing for proton bosons and $\rho= \nu$ standing for neutron bosons, respectively. The microscopic structure
of the model suggests that only two terms are very important. One is
the pairing interaction between identical nucleons, the other is the quadrupole-quadrupole interaction
between non-identical nucleons. The simple standard IBM-2 Hamiltonian \cite{Iachello87,Otsuka92}
is written as
\begin{equation}
\label{Hamito1-1}
\hat{H} = \varepsilon_{d\pi}\hat{n}_{d\pi} + \varepsilon_{d\nu}\hat{n}_{d\nu} +
\kappa_{\pi\nu}\hat{Q}_{\pi} \cdot \hat{Q}_{\nu}\, .
\end{equation}
where $\hat{n}_{d\rho} = d^{\dag}_\rho \cdot\tilde{d_\rho} \, $ and
$\hat{Q}_{\rho} = (s^{\dag}_{\rho} \tilde{d}_{\rho} +
d^{\dag}_{\rho} {s}_{\rho})^{(2)} +
\chi_{\rho}(d^{\dag}_{\rho}\tilde{d}_{\rho})^{(2)} \, $
represent $d$-boson number operator and quadrupole operator, respectively.
The parameter ${\chi_\rho}$ determines the type of the deformation
of the quadrupole operator.  $\varepsilon_{d\rho}$ is the energy of the
\emph{d} bosons relative to the $s$ bosons, $\kappa_{\pi\nu}$ is the strength of the
quadrupole-quadrupole interaction between neutron boson and proton boson.\

The Hamiltonian of Eq.(\ref{Hamito1-1}) has a much richer shape phase structure,
which contains $\textrm{U}_{\pi \nu }(5)$, $\textrm{O}_{\pi \nu }(6)$,
$\textrm{SU}_{\pi \nu }(3)$, $\overline{\textrm{SU}_{\pi \nu }(3)}$, and $\textrm{SU}^{\ast}_{\pi
\nu }(3)$ dynamical symmetries, corresponding to the spherical vibrator, $\gamma$-unstable
rotor, axially symmetric prolate rotor, axially symmetric oblate rotor, and triaxial rotor,
respectively. The shape phase transitions in nuclei can be characterized as the quantum phase
transitions in between the different dynamical symmetries in the IBM \cite{Caprio05,Cejnar09,Zhangyu20141}.
Although the standard IBM-2 Hamiltonian can give us more clear space
of dynamical symmetry, one has to add other physical dominant interactions in order to described
real nuclei more accurately. We use the following IBM-2 Hamiltonian in this paper \cite{Iachello87,Pietralla08,Zhang15},
\begin{equation}
\label{Hamil1-2}
\hat{H} = \varepsilon_{d\pi}\hat{n}_{d\pi} +  \varepsilon_{d\nu}\hat{n}_{d\nu}  +
\kappa_{\pi\nu}\hat{Q}_{\pi} \cdot \hat{Q}_{\nu}
+\omega_{\pi\nu}\hat{L}_{\pi}\cdot\hat{L}_{\nu}+\hat{M}_{\pi\nu} \, ,
\end{equation}
where $\hat{L_\rho}=\sqrt{10}[d^{\dag}_\rho \cdot\tilde{d_\rho}]^{(1)}$ is the angular momentum operator
with a dipole proton-neutron interaction parameter $\omega_{\pi\nu}$, and
$\hat{M}_{\pi\nu} = \lambda_2(s^{\dag}_{\pi}d^{\dag}_{\nu} -
s^{\dag}_{\nu} d^{\dag}_{\pi})^{(2)} \cdot
(s_{\pi}\tilde{d}_{\nu}-s_{\nu}\tilde{d}_{\pi})^{(2)}$\
$+\sum_{k=1,3}\lambda_{k}(d^{\dag}_{\pi}
d^{\dag}_{\nu})^{(k)}\cdot(\tilde{d}_{\pi}\tilde{d}_{\nu})
 ^{(k)}\, $
is the Majorana interaction, the Majorana parameters $\lambda_{k}$ (\emph{k}=1,2,3) represent
the strength of the Majorana interaction.\ If one does not include the Majorana term in
Eq. (\ref {Hamil1-2}), the adopted parameters in the remaining four terms of Eq. (\ref {Hamil1-2})
indicate that the nucleus described by the IBM-2 may locate on the the plane
of  $\textrm{U}_{\pi \nu }(5)$-$\textrm{O}_{\pi \nu }(6)$-$\textrm{SU}^{\ast}_{\pi \nu }(3)$,
because the $\hat{L}_{\pi}\cdot\hat{L}_{\nu}$ only generate the physical rotational
group $\textrm{SO}_{\pi \nu }(3)$ of the IBM-2 \cite{Iachello87,Caprio05,Cejnar09}.
On the other hand, in Ref.\cite{zhangyu20142} the authors proposed a new algebraic model $F(5)$
based on the Euclidean dynamical symmetry in five dimensions ($\rm{Eu(5)}$), which can build a
symmetry intermediate between the $E(5)$ and $X(5)$ symmetries \cite{Zhangyu20141}. However,
the $F(5)$ cannot directly be defined in the IBM-1 or IBM-2 due to the non-compactness of
the $\rm{Eu(5)}$ group \cite{zhangyu20142}.\

The decay properties of the low-lying states, such as the $B(E2)$ transition probabilities,
the magnetic dipole $B(M1)$ transition strengths and the $\rho^2(E0)$ values between lowest
0$^+$ states are the signatures of phase structure. In the IBM-2, the \emph{E}2 transition
strength is given by the following expression
\begin{equation} \label{T-Multipole}
B(E2,J'{\rightarrow}J) =
\frac{1}{2J'+1}|{J\langle}\|\hat{T}{(E2)}\|J'\rangle|^2\, .
\end{equation}
where $J'$ and $J$ are the angular momenta for the initial and final states, respectively.
The $\hat{T}{(E2)}=e_{\pi} \hat{Q}_{\pi} + e_{\nu}\hat{Q}_{\nu}$ is $E$2 operator, where
the operator $Q_\rho$ is the same as in Eq.(\ref{Hamito1-1}). The parameters $e_\pi$ and $e_\nu$
are the effective charges of proton bosons and neutron bosons, respectively. The values of
$e_\nu$ and $e_\pi$ could be taken differently.\

In the proton-neutron interacting boson model, the magnetic dipole $M1$ transition operator
is defined as
\begin{equation} \label{M-Multipole}
\hat{T}{(M1)}=\sqrt{\frac{3}{4\pi}}(g_\nu\hat{L}_\nu+g_\pi\hat{L}_\pi),\
\end{equation}
where the $\hat{L}_\rho$ is the same as in Eq.(\ref{Hamil1-2}). The $g_\pi$ and $g_\nu$
are the effective proton and neutron boson $g$-factors, repectively. Typically, one can
take the values of $g_\pi=1$, $g_\nu=0$ in the calculations \cite{Otsuka92}.\

The $E0$ transition matrix element $\rho$ in the IBM-2 is written as  \cite{Giannatiempo93, Zhang18}
\begin{equation} \label{p0-Operator}
\rho(E0,J'{\rightarrow}J) =\frac{Z}{e R^2}[
\beta_{0\pi}{\langle}J\|\hat{T}_\pi{(E0)}\|J'\rangle+
\beta_{0\nu}{\langle}J\|\hat{T}_\nu{(E0)}\|J'\rangle]\ ,
\end{equation}
where $R=1.2A^{1/3}$fm, $\beta_{0\pi}$,and $\beta_{0\nu}$ are the effective monopole charges of proton
and neutron boson in units of $e$fm$^2$, respectively. The $E$0 operator is expressed as
$\hat{T}{(E0)}=\beta_{0\pi}\hat{T}_\pi{(E0)} + \beta_{0\nu}\hat{T}_\nu{(E0)}
=\beta_{0\pi} \hat{n}_{d\pi} + \beta_{0\nu} \hat{n}_{d\nu}$,
where the $\hat{n}_{d\rho}$ is the same as in Eq.(\ref{Hamito1-1}).\
%

\section{Results and discussion}
\label{s3}
The $^{124}$Te situates at the position between vibrational nucleus and deformed nucleus \cite{Vanhoy04}.
The IBM provides a powerful tool to describe the nuclear shapes and the shape phase transitions. With this approach,
a lot of works have been devoted to describe the properties of the low-lying states of $^{124}$Te isotope.
In IBM-1, the early work regarded the $^{124}$Te as an example of O$(6)$ symmetry, but the energy levels of
yrast states of $^{124}$Te differ from the O(6) limit \cite{Robinson83}.
The further systematic investigations found that $^{124}$Te was close to the U$(5)$
symmetry \cite{Kern95,Pascu10}.
Moreover, recent results of experiments and calculations indicated that $^{124}$Te may posses the $E(5)$
features \cite{Hicks17,Clark04,Sabria16,Ghit08}. In IBM-2, the calculations with and without mixed
configuration suggested that $^{124}$Te is an O$(6)$-like nucleus, but the $B(E2)$ decay pattern is not consistent
with this symmetry \cite{ RIKOVSKA89}. The subsequent study shows that neither O$(6)$ nor U$(5)$ symmetries can
describe the $^{124}$Te well \cite{Warr98}, although the potential-energy surface of $^{124}$Te exhibits an obvious
manifestation of O$(6)$ limit \cite{SubberJf87}. Another recent experiment and IBM calculation support
that $^{124}$Te shows the soft triaxial behavior \cite{Saxena14}. Although the debate on the low-lying structure
of the $^{124}$Te still exists, it is indubitable that all
these investigations indicated that $^{124}$Te is a transitional nucleus between vibration U$(5)$
symmetry and $\gamma$-unstable O$(6)$ symmetry in IBM, and might accompany with somewhat
soft triaxial rotation. \

\begin{figure}
\begin{center}
\includegraphics[width=12cm]{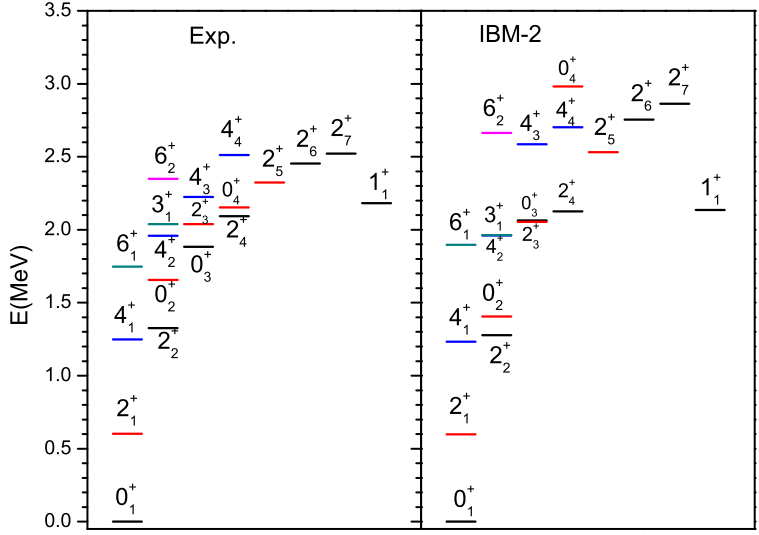}
\caption{
\label{figzhangdl}
Comparison of the calculated and experimental \cite{Hicks17,Egidy06}
low-lying positive-parity levels of $^{124}$Te.}
\end{center}
\end{figure}

To describe the low-lying structure of $^{124}$Te, we select the doubly closed shell
$^{132}_{50}$Sn$_{82}$ as the inert core. There are $N_\pi=1$ particle-like
boson beyond the $Z=50$ major shell and $N_\nu=5$ hole-like neutron bosons from
the $N=82$ shell closure, and the total number of bosons is $N_B=6$.
Considering that the proton boson is particle-like and the neutron bosons are hole-like based
on the different single particle orbitals in $^{124}$Te, we take different relative energies for $d$
neutron and proton boson in this case, $i.e.$, $\varepsilon_{d\pi}\neq\varepsilon_{d\nu}$
just as in Refs.\cite{Dejbakhsh92,zhangding14,zhangmulett16}. To describe the different
type of valence neutrons and protons exhibiting opposite intrinsic quadrupole deformation,
and obtain the $\gamma$-unstable O$(6)$ symmetry, we adopt the $\chi_\pi+\chi_\nu=0$. In order
to introduce the somewhat character of soft triaxial rotation in $^{124}$Te, the value of $\chi_\pi$
and $\chi_\nu$ should not be zero. In principle, the parameter $\varepsilon_{d\rho}$ mainly
leads to vibrational solutions, while $\kappa_{\pi\nu}$ drives the system into a deformed
shape, and the description of the electromagnetic properties is very sensitive to the parameters $\chi_\rho$.
The dipole interaction $\hat{L}_{\pi}\cdot\hat{L}_{\nu}$ can improve the rotational spectrum
and can adjust the position of the 4$^+_1$ relative to 2$^+_2$ state, but it does not affect
the wave vectors \cite{Pietralla98}. The Majoranan operator mainly contributes to the mixed
symmetry states including the scissors mode. All the free parameters in our calculations were
fixed to reproduce the experimental energies and electromagnetic transition strengths of $^{124}$Te.
By fitting to the energy levels of experimental data especially for the 2$^+_1$ state,
we obtain the parameters of IBM-2 in this work as follows:
$\varepsilon_{d\pi}=1.220$MeV, $\varepsilon_{d\nu}=0.710$MeV,
$\kappa_{\pi\nu}=-0.140$MeV, $\chi_\pi=-\chi_\nu=-1.00$, $\omega_{\pi\nu}=-0.055$MeV,
$\lambda_1=-0.700$MeV, $\lambda_2=0.220$MeV, and $\lambda_3=-0.100$MeV. The IBM-2 Hamiltonian
will be numerically diagonalized by using the NPBOS code \cite{Otsuka85}. In Figure \ref{figzhangdl},
we compare the calculated level energies with the experimental results \cite{Hicks17,Egidy06}.
The experimental level energies in Figure \ref{figzhangdl} are separated into different bands
based only on their order of appearance, where only the low-spin positive parity states with
uniquely assigned spin and parity are displayed. \

%
%

Figure \ref{figzhangdl} shows that the theoretical levels are good consistent with
the corresponding experimental ones. In particular, the 6$^+_1$ state is dominated
by two valence protons and is relatively unaffected by the increase of neutrons,
the 6$^+_1$ states remain at a nearly constant excitation energy for even-even
$^{122-130}$ Te isotopes \cite{Hicks17}. By introducing the dipole interaction term
into the Hamiltonian \cite{Nomura11, Nomura16, Zhangmu17}, the structure of the yrast
bands including the 6$^+_1$ level are reproduced by the theoretical predications.
Similar to the 6$^+_1$ state, the observed energy of 4$^+_2$ level is only changed
about 135 keV from $^{122}$Te to $^{130}$Te isotopes. The present calculation reproduces
the experimental energy of the 4$^+_2$ level nicely.
Meanwhile, the present calculated energies of the 2$^+_{3,4,5}$, and 6$^+_2$ levels
agree with the experimental observations. Even though the 0$^+_2$ level is obviously
higher than that is expected for a two-phonon multiple state, the description of the first
two excitation 0$^+$ states in energy sequence is satisfactory. Furthermore, the calculated
energy level of the first scissor mode 1$^+_1$ perfectly reproduces the experimental data.
However, there are some discrepancies between the experimental observations and the theoretical
predications for higher energy states, such as 2$^+_6$, and 2$^+_7$ states,
which is a general feature of this model \cite{Kim96}. For the 0$^+_4$ state, the calculated level
is significantly higher than the experimental one. The IBM-2 with the mixed configuration shows
that 0$^+_4$ is an intruder state in $^{124}$Te and locates outside the IBM-2 space \cite{RIKOVSKA89}.\

\begin{table}[ht]
\begin{center}
\caption{\label{T1}
Experimental $E2$ transition strengths in $^{124}$Te in units of W.u.
are compared with the calculated results.
The experimental data are taken from Refs.
\cite{Garrett18, Hicks17, Ghit08, Rikovska89, Hicks08, Doll00}}.
\footnotesize
\begin{tabular*}{80mm}{c@{\extracolsep{\fill}}cccc}
\hline
{$J^\pi_i\rightarrow$$J^\pi_f$}  &Expt. &IBM-2 \\
\hline
$2^+_1$$\rightarrow$$0^+_1$&31$^5_5$                &  31.02   \\
$4^+_1$$\rightarrow$$2^+_1$&$54.51$                 &  48.06 \\
$2^+_2$$\rightarrow$$2^+_1$&55.5$^{109}_{99}$       &  49.53 \\
$2^+_2$$\rightarrow$$0^+_1$&0.83$^{23}_{16}$        &  0.24  \\
$0^+_2$$\rightarrow$$2^+_1$&20$^{4}_{4}$            &  27.92  \\
$0^+_3$$\rightarrow$$2^+_2$& 50.00                  &  48.51   \\
$4^+_2$$\rightarrow$$2^+_2$&14.1$^{30}_{28}$        &  29.94  \\
$4^+_2$$\rightarrow$$4^+_1$&12.7$^{74}_{59}$        &  27.06 \\
$4^+_2$$\rightarrow$$2^+_1$&4.3$^{9}_{9}   $        &  0.31  \\
$3^+_1$$\rightarrow$$2^+_2$&59$^{10}_{10}   $       &  42.65   \\
$2^+_3$$\rightarrow$$0^+_1$&0.26$^{4}_{4}  $        &  2.08 \\
$2^+_3$$\rightarrow$$2^+_1$&0.025$^{4}_{3}$         &  0.50  \\
$2^+_3$$\rightarrow$$2^+_2$& $\leq2.7^{19}_{15}$    &  4.03   \\
$2^+_3$$\rightarrow$$4^+_1$&0.81$^{104}_{81}$       &  5.38   \\
$2^+_3$$\rightarrow$$3^+_1$&0.26$^{5}_{4}$          &  0.37   \\
$2^+_4$$\rightarrow$$2^+_2$&0.29$^{39}_{33}$        &  1.85   \\
$2^+_4$$\rightarrow$$2^+_1$&1.6$^{3}_{3}$           &  1.39  \\
$2^+_4$$\rightarrow$$0^+_1$&0.053$^{17}_{14}$       &  2.16  \\
$0^+_4$$\rightarrow$$2^+_1$&$<0.5$                  &  0.04  \\
$0^+_4$$\rightarrow$$2^+_2$&$<50$                   &  0.03  \\
$1^+_1$$\rightarrow$$2^+_1$&1.2$^{4}_{3}   $        &  4.99  \\
$1^+_1$$\rightarrow$$2^+_2$&$0.74^{38}_{38}$        &  1.13  \\
$4^+_3$$\rightarrow$$2^+_2$&13.4$^{49}_{39}$        &  1.16 \\
$4^+_3$$\rightarrow$$4^+_1$&6.7$^{23}_{20} $        &  2.89  \\
$4^+_3$$\rightarrow$$2^+_1$&1.9$^{6}_{5}$           &  0.95 \\
$2^+_5$$\rightarrow$$2^+_2$&0.61$^{40}_{30}$        &  1.02   \\
$2^+_5$$\rightarrow$$2^+_1$&0.06$^{7}_{6}$          &  0.15  \\
$2^+_5$$\rightarrow$$0^+_1$&0.078$^{50}_{42}$       & 0.09 \\
\hline
\hline
\end{tabular*}
\end{center}
\end{table}

\begin{table}[ht]
\begin{center}
\begin{tabular}{cccccccccccccccccccccccccccccc}
&Table \ref{T1}(continued)&\\
\hline
{$J^\pi_i\rightarrow$$J^\pi_f$}  &Expt. &IBM-2 \\
\hline
$2^+_6$$\rightarrow$$2^+_2$&1.6$^{5}_{5}$           & 1.67 \\
$2^+_6$$\rightarrow$$4^+_1$&1.7$^{5}_{4}$           & 0.00 \\
$2^+_6$$\rightarrow$$2^+_1$&0.046$^{8}_{6}$         & 0.23 \\
$2^+_6$$\rightarrow$$0^+_1$&0.12$^{3}_{2}$          & 0.01 \\
$4^+_4$$\rightarrow$$4^+_1$&2.6$^{9}_{7}$           & 0.10 \\
$4^+_4$$\rightarrow$$2^+_1$&0.051$^{28}_{22}$       & 0.72 \\
$2^+_7$$\rightarrow$$2^+_1$&4.0$^{4}_{4}$            & 0.56 \\
$2^+_7$$\rightarrow$$2^+_2$&0.20$^{10}_{11}$        & 0.00 \\
$2^+_7$$\rightarrow$$4^+_1$&$<1.6$                  & 1.10 \\
\hline \hline
\end{tabular}
\end{center}
\end{table}

In seeking to investigate the properties of the $E2$ transitions in $^{124}$Te, the effective
charges of proton and neutron bosons were determined to reproduce the experimental $B(E2)$ values. Using
the same manner as in Refs.\cite{Elhami07,zhangmu16} and exactly fitting to the experimental
data of $B(E2,2^+_1\rightarrow0^+_1)=31(5)$ W.u., we obtain $e_\pi=3.780$ $\sqrt{\rm {W.u.}}$, $e_\nu=1.670$
$\sqrt{\rm {W.u.}}$. The theoretical calculation of $B(E2)$ values for $^{124}$Te
in comparison with the available experimental data \cite{Garrett18, Hicks17, Ghit08, Rikovska89, Hicks08, Doll00}
is given in Table \ref{T1}.\

Table \ref{T1} shows that the computed $\emph{B}(E2)$ transition strengths are
in overall agreement with the experimental data, although most cases of the experimental value
have a large uncertainty. In more detail, the typical strongly collective $E2$
transitions with tens of Weisskopf units are described by the theoretical predications
very well, some of them consist with each other within the experimental uncertainty.
It is noteworthy that the experimental $\emph{B}(E2,4^+_1\rightarrow2^+_1)$ transition
strength is $35.9(\pm 17)$ W.u. in Ref.\cite{Hicks17}, which is taken from
Ref.\cite{Saxena14}. In Ref.\cite{Doll00}, the experimental values show that the
lower and upper limits of this transition probability are 27.3 and 163.5 W.u., respectively.
Similar to Ref.\cite{Rikovska89}, we adopt the experimental averaged value 54.51 W.u.
for this transition, the detail can be seen in the Ref.\cite{Mardirosian1984}, one
can find that the experimental value agrees with the present calculated value. At the same
time, the experimental relatively strong $B(E2)$ transitions, which can be comparable
to the experimental transition probability of the first excited state 2$^+_1$ decay
to the ground state with dozens of W.u., are reproduced by the calculated results
perfectly except the $\emph{B}(E2,4^+_3\rightarrow2^+_2)$ transition strength.
Furthermore, the calculated results are in good description of the properties of the
experimental weakly collective $E2$ transitions with about one or even less than one W.u..
The theoretical $E2$ transition strengths from the scissor
mode 1$^+_1$ to the 2$^+_1$ and 2$^+_2$states are in agreement with the corresponding
experimental data, although the theoretical values are a little higher than the
experimental data.\

To identify where the $^{124}$Te can be placed in the $\textrm{U}_{\pi \nu}(5)$
to $\textrm{O}_{\pi \nu}(6)$ transition, we focus on a set of key observables \cite {Casten00},
such as the energy ratios
$R_{4_1/2_1}=E(4^+_1)/E(2^+_1)$,
$R_{2_2/2_1}=E(2^+_2)/E(2^+_1)$,
$R_{0_2/2_1}=E(0^+_2)/E(2^+_1)$,
$R_{0_3/0_2}=E(0^+_3)/E(0^+_2)$,
and the $B(E2)$ ratios $R_{B,42}=B(E2,
4^+_1\rightarrow2^+_1)/B(E2, 2^+_1\rightarrow0^+_1)$,
$R_{B,22}=B(E2, 2^+_2\rightarrow2^+_1)/B(E2, 2^+_1\rightarrow0^+_1)$
and $R_{B,02}=B(E2, 0^+_2\rightarrow2^+_1)/B(E2,
2^+_1\rightarrow0^+_1)$, which reflect the characteristics of the nucleus behavior
at critical point of the phase transition from $\textrm{U}_{\pi \nu}(5)$ to
$\textrm{O}_{\pi \nu}(6)$ in IBM-2 space \cite{Caprio05}, and are the most
crucial nuclear structure indicators \cite{Casten88}. Some of the
indicators can even distinguish the first order quantum phase
transition from the second order quantum phase transition \cite{IZ04, Liu07, Bonatsos081}.
The values of these key observables at the critical point of the phase transition from
$\textrm{U}_{\pi \nu}(5)$ to $\textrm{O}_{\pi \nu}(6)$ for the infinite numbers of bosons \cite{Iachello00},
as well as the experimental data and the calculated results in $^{124}$Te are listed
in Table \ref{T2} for comparison.\
\begin{table}[ht]
\begin{center}
\caption{ \label{T2}Comparison of the key observables of the states at the critical point of
the $\textrm{U}_{\pi\nu}(5)$--$\textrm{O}_{\pi \nu}(6)$ transition (taken from
Ref.\cite{Iachello00} and denoted as CPST), the experimental data of $^{124}$Te
(taken from Refs.\cite{Garrett18,Hicks17, Rikovska89, Hicks08, Doll00} and labeled as $^{124}$Te)
and the calculated results (labeled as IBM-2). }
\footnotesize
\begin{tabular*}{80mm}{c@{\extracolsep{\fill}}cccccccc}
\hline
               &CPST &   $^{124}$Te&  IBM-2      \\
\hline
$R_{4_1/2_1}$& 2.20& 2.07& 2.06\\
$R_{2_2/2_1}$& 2.20& 2.20& 2.13\\
$R_{0_2/2_1}$& 3.03& 2.75& 2.35\\
$R_{0_3/0_2}$& 1.18& 1.14& 1.46\\
$R_{B,42}$   & 1.68& 1.74& 1.55\\
$R_{B,22}$   & 1.68& 1.79& 1.60\\
$R_{B,02}$   & 0.86& 0.65& 0.90\\
\hline
\hline
\end{tabular*}
\end{center}
\end{table}

Table \ref{T2} shows that the overall agreement is  well. On the energy ratios,
the experimental ratio $R_{2_2/2_1}$ is almost exactly reproduced to the
the critical point of the $\textrm{U}_{\pi \nu}(5)$--$\textrm{O}_{\pi \nu}(6)$ transition.
The experimental ratio $R_{0_3/0_2}$ is very close to the value of CPST.
Meanwhile, the experimental energy of the $0^+_2$ state relative to $E(2^+_1)$
satisfactorily matches the predicated value of CPST. The experimental
$R_{4_1/2_1}$ lies between U$_{\pi\nu}(5)$ ($R_{2_2/2_1}=2.00$) and
$\textrm{O}_{\pi \nu}(6)$ ($R_{2_2/2_1}=2.50$) limits, although it deviates slightly
from the predicated one. On the $B(E2)$ ratios, the experimental $R_{B,42}$ is well
reproduced by the theoretical predication and clearly points to a structure intermediate
between U$_{\pi\nu}(5)$ ($R_{B,42}=2.00$) and $\textrm{O}_{\pi \nu}(6)$($R_{B,42}=10/7$).
Meanwhile, the experimental value of $R_{B,22}$ approaches the predicted value of the critical
point theory, too. In particular, the information of $ R_{B,02}$ is a signature for
identifying the $\textrm{U}_{\pi \nu}(5)$ from the $\textrm{O}_{\pi \nu}(6)$ symmetry.
In $\textrm{U}_{\pi \nu}(5)$ limit the $B(E2, 0^+_2\rightarrow2^+_1)$ transition strength is
2 times larger than $(B(E2, 2^+_1\rightarrow0^+_1)$. In $\textrm{O}_{\pi \nu}(6)$ limit
the $B(E2, 0^+_2\rightarrow2^+_1)$ transition is forbidden \cite{Stachel82}.
Here, both the experimental and theoretical values of the $R_{B,02}$ can correctly reflect
the nature of the $B(E2, 0^+_2\rightarrow2^+_1)$ transition, although the calculated value
overestimates about 1.5 times than experimental one. Therefore, all the available experimental
information on the key observables for $^{124}$Te is in good agreement with the predictions
of critical point of the $\textrm{U}_{\pi \nu}(5)$--$\textrm{O}_{\pi \nu}(6)$ transition.
Meanwhile, Table \ref{T2} also shows that the present calculation of the characteristic
feature of $^{124}$Te is remarkable. However, the $R_{0_2/2_1}$ of CPST seems larger
than the IBM's in Table \ref{T2}, the reason is that the calculation of CPST is given for the
infinite$-N_B$ limit \cite{Iachello00,Casten00}. For the finite number of bosons, Ref.\cite{Caprio05}
gives that the $R_{0_2/2_1}$ of a nucleus at the critical point of the phase transition
from $\textrm{U}_{\pi \nu}(5)$ to $\textrm{O}_{\pi \nu}(6)$ with $N_{\pi} = N_{\nu} =5$
is 2.48, which agrees the result of present calculation. From the above discussion, we conclude
that the $^{124}$Te may be a nucleus at the critical point of
the $\textrm{U}_{\pi \nu}(5)$--$\textrm{O}_{\pi \nu}(6)$ transition. Consequently,
the $0^+_2$ state is the possible lowest state of the first excited family of intrinsic
levels predicted by the critical point symmetry. \

An advantage of IBM-2 over IBM-1 is that the former can study the influence of the critical point
symmetry on magnetic dipole transitions between the low-lying states \cite{Pietralla08}.
In IBM-2, the pure $\textrm{O}_{\pi\nu}(6)$ symmetry can be obtained only
with $\chi_\pi=\chi_\nu=0$. In this case, there does not exist any asymmetry between $\chi_\pi$ and $\chi_\nu$.
It means that the mixed symmetry components could not mix into the low-lying states. Therefore, the $M1$
transitions between these states are forbidden in $\textrm{O}_{\pi\nu}(6)$ limit \cite{Otsuka92}.
The theoretical predictions show that the $B(M1)$ transition strengths among the low-lying
states almost vanish,
when a nucleus is at the critical point of the $\textrm{U}_{\pi \nu}(5)$--$\textrm{O}_{\pi \nu}(6)$
transition \cite{Caprio05}. For the scissors mode state 1$^+_1$, the $B(M1,1^+_1\rightarrow 0^+_1)$
transition strength is $3N_\pi N_\nu(g_\nu-g_\pi)^2/{4\pi(2N_B+1)}$ for
$\textrm{O}_{\pi \nu}(6)$ limit, but it is forbidden for $\textrm{U}_{\pi\nu}(5)$
limit. The $B(M1,1^+_1\rightarrow 0^+_2)$ value is
$3N_\pi N_\nu(g_\nu-g_\pi)^2/{\pi N_B(N_B-1)}$ for $\textrm{U}_{\pi\nu}(5)$ limit,
whereas vanish for $\textrm{O}_{\pi\nu}(6)$ limit \cite{Pietralla08, Backer86, Sevrin87}.
At the critical point of $\textrm{U}_{\pi \nu}(5)$--$\textrm{O}_{\pi \nu}(6)$ transition,
the 1$^+_1$ state has an allowed $M1$ decays to first and second $0^+$ states.
In the case of $^{124}$Te, the $B(M1,1^+_1\rightarrow 0^+_1)$
is 0.10 $\mu^2_N$ for $\textrm{O}_{\pi \nu}(6)$ limit, and $B(M1,1^+_1\rightarrow 0^+_2)$
is 0.50 $\mu^2_N$ for $\textrm{U}_{\pi\nu}(5)$ limit
with the typical values $g_\pi=1$, $g_\nu=0$.
Indeed, only a few $M1$ transitions with very small absolute values of $B(M1)$
among the low-lying states in $^{124}$Te have been measured \cite{Hicks08,Georgii95}.
By using the Eq.(\ref{M-Multipole}) and also taking
$g_\pi=1$ and $g_\nu=0$, the calculated $M1$ transition probabilities, as well as the
corresponding available experimental data for the scissors mode and the low-lying states
in $^{124}$Te are given in Table \ref{T3}.\

As can be seen from Table \ref{T3}, all the experimental $B(M1)$ transition strengths
between the lowest 2$^+$ states are of the order of about 0.01 $\mu^2_N$ or even less,
which is far from the typical $B(M1)$ value of the mixed symmetry state.
This demonstrates that the observed $M1$ transitions are in qualitative
agreement with the predictions of the critical point symmetry of the
$\textrm{U}_{\pi \nu}(5)$--$\textrm{O}_{\pi \nu}(6)$ transition. At the same time,
Table \ref{T3} shows that the
calculated results reproduce the characteristics of the $M1$ transitions among low-lying
states very fairly. In particular, the calculated $B(M1,1^+_1\rightarrow 0^+_1)$ strength
is close to the experimental one, which is
comparable the predicted value of the critical
point between the $\textrm{U}_{\pi \nu}(5)$ and $\textrm{O}_{\pi \nu}(6)$ limits.
This indicates that both the experimental and the calculated results of the
$B(M1,1^+_1\rightarrow 0^+_1)$ transition reflect the property of the nucleus at the
critical point of the $\textrm{U}_{\pi \nu}(5)$--$\textrm{O}_{\pi \nu}(6)$ transition.
Unfortunately, the $B(M1,1^+_1\rightarrow 0^+_2)$ strength in $^{124}$Te has not been measured,
the calculated value for this transition is 0.035 $\mu^2_N$, locating between the
$\textrm{U}_{\pi \nu}(5)$ and $\textrm{O}_{\pi \nu}(6)$ transition similar
to the case of $B(M1,1^+_1\rightarrow 0^+_1)$ transition. All these results indicate that both the calculated
and experimental $M1$ transitions are consistent with the characteristic behavior of the $^{124}$Te
nucleus at the critical point of the $\textrm{U}_{\pi \nu}(5)$--$\textrm{O}_{\pi \nu}(6)$
transition.\
\begin{table}[ht]
\begin{center}
\caption{ \label{T3} Comparison of the calculated $M1$ transition strengths
(in units of $\mu^2_N$) and
the experimental data for $^{124}$Te. The experimental data are taken from
Refs.\cite{Hicks08, Doll00,Hicks12, Guliyev02}.}
\footnotesize
\begin{tabular*}{80mm}{c@{\extracolsep{\fill}}ccccccccccccc}
\hline
Transition &                          Expt. & IBM-2 \\
\hline
$2^+_2$$\rightarrow$$2^+_1$ & 0.0003&  0.006\\
$2^+_3$$\rightarrow$$2^+_2$ & 0.004 &  0.025\\
$2^+_3$$\rightarrow$$2^+_1$ & 0.013 &  0.072\\
$2^+_4$$\rightarrow$$2^+_2$ & 0.001 &  0.031\\
$2^+_4$$\rightarrow$$2^+_1$ & 0.004 &  0.075\\
$1^+_1$$\rightarrow$$0^+_1$ & 0.04  &  0.029\\
$1^+_1$$\rightarrow$$0^+_2$ &       &  0.034\\
\hline
\hline
\end{tabular*}
\end{center}
\end{table}

However, some deviations between the theoretical
and experimental $M1$ transition strengths also are displayed in Table \ref{T3}, all the calculated $M1$
values are a little larger than the corresponding experimental data.
In IBM-2, $\chi_\pi=-\chi_\nu\neq0$ leads to spectrum having many of the O(6) features.
In addition, some mixed symmetry components from the asymmetry of $\chi_\pi$ and $\chi_\nu$ mix
into the low-lying states beside the pure mixed symmetry states
at higher energies \cite{Otsuka92,zhangmusci18}. On the other hand,
$\chi_\pi=-\chi_\nu\neq0$ makes the nucleus have some features of the
$\textrm{SU}^\ast_{\pi \nu}(3)$ symmetry, because the proton and neutron have
opposite types of deformation \cite{Cejnar09}.
The calculated $B(M1)$ values among the lowest states are systematically slightly larger than the
experimental data, which implies that the description of $^{124}$Te with the character
of the $\textrm{SU}^\ast_{\pi \nu}(3)$ is slightly rather than the actual nucleus.
Furthermore, the small $B(M1)$ values among the low-lying states in $^{124}$Te also
indicate that $^{124}$Te may contain some components of $\textrm{SU}^\ast_{\pi \nu}(3)$ symmetry.
Therefore, the $^{124}$Te may be the nucleus at the critical point of
the $\textrm{U}_{\pi \nu}(5)$--$\textrm{O}_{\pi \nu}(6)$ transition and with
somewhat $\textrm{SU}^\ast_{\pi \nu}(3)$ symmetry.\

The electric monopole transition strengths between 0$^+$ states can be considered as
a signature of shape coexistence \cite{Heyde11}, quantum phase transitions \cite{Nomura17},
and can reflect the properties of $E(5)$ dynamical symmetry \cite{Clark04}. In fact,
the $\rho^2(E0,0^+_2\rightarrow0^+_1)$ value of $^{124}$Te has been measured \cite{T05}.
But the $\rho^2(E0)$ value on the first excited 0$^+$ state in $^{124}$Te
is a puzzle for a long time as mentioned in the introduction \cite{Garrett16}.
In IBM, the $E0$ transitions do not occur in the U(5) dynamical limit,
because $E0$ operator is proportional to $\hat{n}_d$ \cite{Leviatan16}.
While in O(6) limit, the selection rules require the $E0$ transition from the 0$^+_2$ to
ground state is forbidden \cite{Wood99}. For the predicted $E(5)$ symmetry, the 0$^+_2$ level, which is
the lowest member (zero phonon) of the first excited family, should have an allowed $E0$ branch
to the 0$^+_1$ state. However, so far the studies of the behavior of $E(5)$ symmetry in
$^{124}$Te have not provide a detailed analysis of the properties of $E0$ transitions.
In IBM-2 space, all the symmetrical states have correspondence with the IBM-1 states, therefore, IBM-2 subsumes
the critical point of second order transition of IBM-1\cite{Caprio05, Pietralla08}.
The critical point symmetry of $\textrm{U}_{\pi \nu}(5)$--$\textrm{O}_{\pi \nu}(6)$
bears considerably resemblance to the predictions of the ${E}(5)$ symmetry \cite{Caprio05}.
Subsequently, the $0^+_2$ level should also have an allowed $E0$ decay to the ground state in
this prediction of the critical point symmetry.
We take the values of parameters $\beta_{0\nu}$ and $\beta_{0\pi}$ as in Ref.\cite{Giannatiempo91},
namely, $\beta_{0\nu}=0.25$ and $\beta_{0\pi}=0.10$ $e$fm$^2$. The calculated
$\rho^2(E0,0^+_2\rightarrow0^+_1)\times10^3$ value is $11.00$, which is consistent with the
experimental data within the experimental uncertainty. The theoretical and experimental $E0$ transitions
prove that the 0$^+_2$ state may be interpreted as the lowest state of the first excited family of
intrinsic levels in the critical point symmetry of second order
$\textrm{U}_{\pi \nu}(5)$--$\textrm{O}_{\pi \nu}(6)$ transition.\

It is well known that the low-lying yrast states in other Te isotopes such as $^{112,114}$Te
show the similar behavior, i.e., typical vibrational like spectra but the other properties
deviate from vibrator \cite{Saxena14,Moller2005,Qi2016,Doncel2017}. All these Te isotopes locate
around the midshell $^{118}$Te. The large-scale shell-model calculations indicate that the
behavior of the $B(E2)$ values of these nuclei is related to the competition between the seniority
coupling and the neutron-proton correlations. From the point of IBM model, $^{112}$Te is below
the midshell $^{118}$Te, it has $N_\nu=5$ bosons beyond $N=50$ major shell. Both $^{112}$Te
and $^{124}$Te have the same numbers of bosons with $N_\pi=1$ and $N_\nu=5$, therefore, they
show the similar properties of spectrum. As for $^{114}$Te, the measured ratio of
$B(E2,4^+_1\rightarrow 2^+_1)$ to $B(E2,2^+_1\rightarrow 0^+_1)$ is smaller than one, which
is very unusual deformation, no theoretical models can give a satisfied description \cite{Doncel2017}.\

\section{Conclusion}\
\label{s4}
In summary, the low-lying structure of the $^{124}$Te has been investigated
within the framework of IBM-2. The calculated low-lying energy levels are good consistent with
the corresponding experimental data. In particular, the 6$^+_1$, 4$^+_2$ levels,
which are remaining at a nearly constant excitation energy from $^{122}$Te to $^{130}$Te isotopes,
have been reproduced by calculated results very satisfactorily. Meanwhile, the description of the first
two excitated 0$^+$ states in energy sequence is very well, although the 0$^+_2$ level is obviously
higher than that is expected for a two-phonon multiple state. Furthermore, the calculated
energy level of the first scissor mode 1$^+_1$ perfectly reproduces the experimental data.\

The calculations show that the computed $\emph{B}(E2)$ transition strengths are in
the overall agreement with the experimental data. The observed typical strongly collective $E2$
transitions with tens of Weisskopf units, and the most of the experimental relatively strong
$B(E2)$ transitions comparable to the experimental $\emph{B}(E2,2^+_1\rightarrow0^+_1)$
transition probability with dozens of W.u., are reproduced by the theoretical predications
nicely. Furthermore, the calculated results provide the perfect description of the properties of
the experimentally weakly collective $E2$ transition with about one or even less than one W.u..
Particularly, the theoretical $E2$ transitions from the scissor mode 1$^+_1$
to the 2$^+_1$ and 2$^+_2$ states are in good agreement with the corresponding experimental data,
although the theoretical values are slightly higher than the experimental data.\

By comparing the key observables of the states at the critical point of the
U$_{\pi\nu}(5)$--$\textrm{O}_{\pi \nu}(6)$ transition with the experimental data and
IBM-2 calculations, we show that both the experimental data and the theoretical results
of the $^{124}$Te agree with the predicted properties of the states at
the critical point of the phase transition from $\textrm{U}_{\pi \nu}(5)$ to
$\textrm{O}_{\pi \nu}(6)$ very well. All these information on the key observables indicates
that the $^{124}$Te may be a nucleus at the critical point of
the $\textrm{U}_{\pi \nu}(5)$--$\textrm{O}_{\pi \nu}(6)$ transition. Consequently,
the $0^+_2$ state is the possible lowest state of the first excited family of intrinsic
levels predicted by the critical point symmetry.\

At the same time, the characteristics of the $B(M1)$ transition strengths among the
low-lying states are consistent with the predications of the critical
point of the transition from $\textrm{U}_{\pi \nu}(5)$ to $\textrm{O}_{\pi \nu}(6)$
well, and the overall agreement between the experiment and the theory for
characteristics of the $B(M1)$ transition strengths is perfect. Especially, both
the experimental and theoretical $B(M1,1^+_1\rightarrow 0^+_1)$ values are located
between the $\textrm{U}_{\pi \nu}(5)$ and $\textrm{O}_{\pi \nu}(6)$ limits.
By analyzing the calculated results and the structure parameters of quadrupole
operator in this work, all these quantities support that the $^{124}$Te may be the nucleus at the
critical point of the $\textrm{U}_{\pi \nu}(5)$--$\textrm{O}_{\pi \nu}(6)$
transition and with somewhat feature of $\textrm{SU}^\ast_{\pi \nu}(3)$ symmetry.\

The calculated $\rho^2(E0,0^+_2\rightarrow0^+_1)\times10^3$ value also reproduced
exactly the observed data. Both the theoretical and experimental properties of
$E0$ transition indicate that the 0$^+_2$ state in $^{124}$Te may be interpreted as the lowest
state of the first excited family of intrinsic levels at the critical point
of second order $\textrm{U}_{\pi \nu}(5)$--$\textrm{O}_{\pi \nu}(6)$ transition. Therefore, we
conclude that the $^{124}$Te is a possible nucleus at the critical point of
the second order phase transition from vibration to unstable rotation, and such a critical point contains
somewhat triaxial rotation. The 0$^+_2$ state in $^{124}$Te can be interpreted as the
lowest state of the first excited family of intrinsic levels in the critical point symmetry.
However, it should be mentioned that a few of the measured $B(E2)$ and $B(M1)$
values have a large uncertainty and the experimental data of $B(M1,1^+_1\rightarrow 0^+_2)$
is still missing. More investigations on the experiment and the theory are needed to focus
on these aspects.\

\begin{acknowledgments}

We are grateful to Prof. CaiWan Shen, and Dr. XiaoBao Wang of Huzhou University,
Prof. YuXin Liu of Peking University, and Prof. GuiLu Long of Tsinghua University for their
helpful discussions and good suggestions.  This work is supported by the National Natural Science Foundation of China
under grant numbers 11475062, 11747312 and 11147148.

\end{acknowledgments}


\vspace{10mm}

\vspace{-1mm}
\centerline{\rule{80mm}{0.1pt}}
\vspace{2mm}



\clearpage

\end{document}